\documentclass[a4paper,11pt]{article}
\usepackage{pos}
\usepackage{physics}
\title{Recent results on the $\Lambda \rightarrow p\ell \bar{\nu}_\ell$ semileptonic decay}

\author[a]{Simone Bacchio}
\author*[a,b]{Andreas Konstantinou}

\affiliation[a]{Computation-based Science and Technology Research Center,
The Cyprus Institute, 20 Kavafi Str., Nicosia 2121, Cyprus}

\affiliation[b]{Department of Physics, University of Cyprus, P.O. Box 20537, 1678 Nicosia, Cyprus}

\emailAdd{s.bacchio@cyi.ac.cy}
\emailAdd{an.konstantinou@cyi.ac.cy}

\abstract{

We present a lattice-QCD determination of the $\Lambda \to p$ vector and axial-vector form factors, providing theoretical input for studies of the semileptonic decay $\Lambda \to p\ell\bar{\nu}_\ell$.
The calculation is carried out on a single gauge ensemble with physical light, strange, and charm quark masses and delivers a precise determination of the complete set of transition form factors, including second-class contributions.

Using these form factors, we compute decay rates for both the electronic and muonic channels, as well as their ratio, which offers a sensitive test of lepton-flavor universality and possible non-standard scalar or tensor interactions.

This decay mode provides a theoretically well-controlled avenue for extracting the CKM matrix element $|V_{us}|$ from the baryon sector.

Our estimate of $|V_{us}|$ is obtained by combining our recent lattice-QCD results \cite{Bacchio:2025auj} with recent measurements of the relevant branching fraction reported by BESIII \cite{BESIII:2025hgj} and LHCb \cite{LHCb:2025wld}.

}

\FullConference{The 42nd International Symposium on Lattice Field Theory (LATTICE2025)\\
2-8 November 2025\\
Tata Institute of Fundamental Research, Mumbai, India\\}

\let\OLDthebibliography\thebibliography
\renewcommand\thebibliography[1]{
  \OLDthebibliography{#1}
  \setlength{\parskip}{1.4pt}
  \setlength{\itemsep}{0.6pt plus 0.5ex}
}

\begin{document}

\renewcommand*{\bibfont}{\footnotesize}

\setlength{\parskip}{0pt}
\setlength{\jot}{0pt}

\maketitle

\section{Introduction}
Semileptonic hyperon decays (SHD) provide a valuable probe of the interplay between weak and strong interactions in the baryon sector. The decay $\Lambda \to p \ell \bar{\nu}_{\ell}$ proceeds via an $s {\to} u$ transition mediated by a virtual $W$ boson and offers a complementary path for determining the Cabibbo–Kobayashi–Maskawa (CKM) matrix element $|V_{us}|$. Such a determination can be obtained by combining first-principles lattice-QCD calculations of the relevant hadronic matrix elements with experimental measurements of the decay rate.

Among the experimentally accessible SHDs, the channel $\Lambda \to p e \bar{\nu}_{e}$ is especially favorable for phenomenological analyses, as its branching fraction is known with percent-level precision, exceeding the accuracy currently achieved for other baryonic modes~\cite{ParticleDataGroup:2024cfk}. This makes it a particularly promising candidate for an independent extraction of $|V_{us}|$.

Current determinations of $|V_{us}|$ exhibit a persistent tension between values obtained from kaon semileptonic decays~\cite{Moulson:2017ive,FlavourLatticeAveragingGroupFLAG:2024oxs}, hadronic $\tau$ decays~\cite{ExtendedTwistedMass:2024myu}, and constraints from first-row CKM unitarity~\cite{ParticleDataGroup:2024cfk}. In this context, ongoing and planned improvements in the experimental study of $\Lambda$ SHD—including recent measurements from LHCb~\cite{LHCb:2025wld} and BESIII~\cite{BESIII:2025hgj}, as well as projected sensitivities at STCF~\cite{Zhou:2025skh} and FAIR~\cite{Messchendorp:2025men}—highlight the growing importance of independent, theoretically controlled determinations of $|V_{us}|$ from $\Lambda \to p \ell \bar{\nu}_{\ell}$ decays. Achieving this goal requires precise knowledge of the corresponding hadronic matrix elements, in particular the vector and axial form factors (FFs).

In this work we combine recent experimental measurements with our lattice-QCD determination of the FFs~\cite{Bacchio:2025auj} to further investigate the extraction of $|V_{us}|$. The lattice calculation is based on a single $N_f{=}2{+}1{+}1$ gauge ensemble generated at physical light, strange, and charm quark masses by the Extended Twisted Mass Collaboration (ETMC), with the lattice spacing and pion mass determined as described in Ref.~\cite{ExtendedTwistedMass:2022jpw}.

\section{Theoretical Formalism}

The semileptonic decay rate for $\Lambda \to p\ell\bar{\nu}_\ell$ is obtained
from the squared matrix element after integrating over the three-body
phase space,
\begin{equation}
\Gamma =
\frac{G_F^2 |V_{us}|^2}{192\pi^3 m_\Lambda^3}
\int_{m_\ell^2}^{q^2_{\rm max}} dq^2\,
\sqrt{\lambda(m_\Lambda^2,m_p^2,q^2)}\,
L^{\mu\nu}(q^2)\,H_{\mu\nu}(q^2),
\label{eq:GammaMaster}
\end{equation}
where $G_F$ is the fermi constant, $q^2=(p_\Lambda-p_p)^2$ and
$\lambda(a,b,c)=a^2+b^2+c^2-2(ab+bc+ca)$ is the Källén function.
The tensors $L^{\mu\nu}$ and $H_{\mu\nu}$ denote the leptonic and hadronic
contributions, respectively, and are defined as
\begin{equation}
  L^{\mu\nu} \equiv
\left(1-\frac{m_\ell^2}{q^2}\right)^2
\left[
\frac{q^\mu q^\nu}{q^2}
\left(1+\frac{m_\ell^2}{2q^2}\right)
-g^{\mu\nu}
\right] \qq{and }
H_{\mu\nu} \equiv
\sum_{s_p,s_\Lambda}
M_\mu^{\text{V--A}}
\left(M_\nu^{\text{V--A}}\right)^{*}\,,
\end{equation} 
where $g_{\mu\nu}=\text{diag}(+,-,-,-)$ is the Minkowski metric and
$s_p,s_\Lambda$ denote the proton and $\Lambda$ spin indices.
The hadronic matrix element is given by $M_\mu^{\text{V--A}}
\equiv
\langle p(p_p,s_p)|
\bar{u}\,\Gamma_\mu^{\text{V--A}} s
|\Lambda(p_\Lambda,s_\Lambda)\rangle $.

The hadronic matrix element is parameterized in terms of the Weinberg
form factors,
\begin{equation}
\langle p | \bar{u}\gamma_\mu s | \Lambda \rangle =
\bar{u}_p
\left[
\gamma_\mu F_1(q^2)
- i \sigma_{\mu\nu}\frac{q^\nu}{m_\Lambda} F_2(q^2)
+ \frac{q_\mu}{m_\Lambda} F_3(q^2)
\right] u_\Lambda\, , 
\end{equation}
\begin{equation}
\langle p | \bar{u}\gamma_\mu\gamma_5 s | \Lambda \rangle =
\bar{u}_p
\left[
\gamma_\mu G_1(q^2)
- i \sigma_{\mu\nu}\frac{q^\nu}{m_\Lambda} G_2(q^2)
+ \frac{q_\mu}{m_\Lambda} G_3(q^2)
\right]\gamma_5 u_\Lambda \,,
\end{equation}
which fully determine the hadronic tensor $H_{\mu\nu}$.
These form factors can be computed non-perturbatively in lattice QCD and
provide the essential theoretical input for decay-rate predictions.

\subsection{Three-point functions and ground-state dominance}

The $\Lambda \to p$ matrix elements are extracted from ratios of baryonic
two- and three-point correlation functions. The three-point function is defined as
\begin{multline}\label{eq:3pt}
C^\text{V-A}_{\mu\nu}(\vec{p}_N,\vec{p}_\Lambda,t_{\rm s},t_{\rm ins})= \sum_{\vec{x}_{\rm ins},\vec{x}_s} e^{i\vec{x}_{\rm ins}\cdot (\vec{p}_\Lambda-\vec{p}_N)-i\vec{x}_{s}\cdot \vec{p}_\Lambda} Tr\left[P_\nu\langle \chi_\Lambda(x_s) \bar{u}(x_{\rm ins})\,\varGamma^\text{V-A}_\mu s(x_{\rm ins})\bar{\chi}_N(0)\rangle \right]
\end{multline}
where $\chi_p$ and $\chi_\Lambda$ are interpolating fields for the proton and
$\Lambda$ baryon, respectively, $P_\nu$ is a projector, which acts on spin indices, and $\varGamma^\text{V-A}_\mu$ denotes the local
vector - axial-vector current.
The corresponding two-point functions are constructed in the standard way, $C_B(\vec{p}_B,t_{\rm s})=
\sum_{\vec{x}_s} Tr\left[P_0\langle \chi_B(x_s) \bar{\chi}_B(0)\rangle e^{-i\vec{x}_s\cdot \vec{p}_B}\right]$ for $B\in\{N,\Lambda\}$.
Matrix elements are obtained from suitable ratios that cancel unknown overlap
factors and isolate the desired ground-state contribution in the limit of large
source--sink separation.

Controlling excited-state contamination is essential for a reliable extraction
of ground-state matrix elements. In our work, we adopted analysis strategies
that have been successfully applied in recent high-precision studies of nucleon
structure~\cite{Alexandrou:2024ozj,Alexandrou:2023qbg,Alexandrou:2025vto}.
Such treatments are particularly important for nucleon axial-current matrix
elements, where excited-state effects are known to be enhanced by sizable
overlaps with low-lying $\pi N$ states~\cite{Bar:2018xyi,Jang:2019vkm}.

For the strangeness-changing matrix elements considered here, the situation is
qualitatively different. The current couples to states containing a kaon, such
that the lowest multi-hadron state contributing on the nucleon side is $K N$
rather than $\pi N$. As a result, the relevant energy gap is approximately twice
as large as in the nucleon case, leading to a substantially stronger
exponential suppression of excited-state contributions. This behavior is
reflected in the stability of the extracted form factors as the source--sink
separation increases. Residual effects are further
controlled through a systematic study of source--sink separations and
multi-state fit analyses.

\subsection{Parameterization of the $q^2$ dependence}

The relevant momentum-transfer region for the
$\Lambda \to p$ semileptonic decay lies entirely at low values of $q^2$,
approximately within $[0,0.03]~\mathrm{GeV}^2$.
In this region, a leading-order expansion in $q^2$ is sufficient to
determine the decay rate with sub-percent precision.
We therefore define the axial-vector charges, radii and the low-$q^2$ expansion as
\begin{equation}
g_i = G_i(0),
\qquad
\langle r^2_{G_i} \rangle
= \frac{6}{g_i}
\frac{\partial G_i(q^2)}{\partial q^2}
\Big|_{q^2=0}, \qquad G_i(q^2)
=
g_i
\left(1+\frac{\langle r^2_{G_i} \rangle}{6} q^2\right)
+\mathcal{O}(q^4).
\end{equation}
and analogously for the vector form factors by replacing
$G \leftrightarrow F$ and $g \leftrightarrow f$.

In contrast, the lattice data cover a broader kinematic range, including
negative $q^2$ values and extending up to $q^2_{\max}$ for several form
factors.
To describe the full $q^2$ dependence, we employ the model-independent
$z$-expansion~\cite{Hill:2010yb}.

\subsection{From Form Factors to Decay Rates}

In Eq.~\eqref{eq:GammaMaster}, the contraction of the leptonic and hadronic
tensors can be performed using the completeness relation for the polarization
four-vectors~\cite{Korner:1989qb} $\sum_{m,m'=\pm,0,t}
\varepsilon^{\mu}(m)\,\varepsilon^{\nu*}(m')\, g_{mm'}
= g^{\mu\nu}$, which leads to the semi-covariant representation
\begin{equation}
L_{\mu\nu} H^{\mu\nu}
= \sum_{m,n,m',n'}
\left[ L^{\mu'\nu'} \varepsilon_{\mu'}(m)
       \varepsilon^{*}_{\mu}(m') \right]
\left[ H^{\mu\nu}\varepsilon^{*}_{\nu}(n)
       \varepsilon_{\nu'}(n') \right]g_{mm'}\, g_{nn'} .
\end{equation}
So, $L_{\mu\nu} H^{\mu\nu}= \sum_{n,n'}L_{nn'}\, H_{nn'}$, where $L_{nn'}$ and $H_{nn'}$ denote the leptonic and hadronic helicity
tensors, respectively. The hadronic helicity amplitudes are written as $H_{nn'} = H^{V}_{nn'} + H^{A}_{nn'}$, where the vector and axial-vector contributions are expressed in terms of the
Weinberg form factors and are given explicitly in
Eqs.~(9) and (10) of Ref.~\cite{Kadeer:2005aq}.

After integration over the lepton angle $\theta$, the differential decay rate
takes the form
\begin{align}\label{eq:difrat}
&\frac{d\Gamma}{dq^2} =\frac{G_F^2 |V_{us}|^2 \sqrt{s_+ s_-}}{192 \pi^3 m_{\Lambda}^3}\left( 1 - \frac{m_\ell^2}{q^2} \right)^2 
\Bigg\{\frac{3 m_\ell^2}{2q^2}\Big( s_- [(m_{\Lambda} + m_N) G_0]^2 + s_+ [(m_{\Lambda} - m_N) F_0]^2 \Big)+\\ &\frac{m_\ell^2 + 2 q^2 }{2q^2}\Big( s_+ [(m_{\Lambda} - m_N) G_+]^2 + s_- [(m_{\Lambda} + m_N) F_+]^2 \Big)
+\left( m_\ell^2 + 2 q^2 \right)\Big( s_+ [G_\perp]^2 + s_- [F_\perp]^2 \Big)\Bigg\}\,\nonumber
\end{align}
with $s_{\pm}=(m_\Lambda\pm m_N)^2-q^2$, and the differential decay rate is expressed with respect to the helicity form factors which are related to the Weinberg form factros e.g. as in Eqs. (21) - (26) of Ref.~\cite{Bacchio:2025auj}.

In Ref.~\cite{Bacchio:2025auj}, the NLO and NNLO expansions of the electronic decay rate in powers of $\delta=(m_\Lambda-m_p)/m_\Lambda$ are compared to the fully non-perturbative numerical integration of Eq.~\eqref{eq:difrat} employed in the present work.
It is found that neglecting the full $q^2$ dependence of the form factors
induces a deviation of $3.6(3)\%$ in the electron-mode decay rate relative to the fully integrated result.
Including next-to-next-to-leading-order corrections, in which the $q^2$
dependence enters through the form-factor radii, reduces this discrepancy to approximately $-1\%$. These differences demonstrate that a fully non-perturbative treatment of the complete $q^2$ dependence is required for precision studies of $\Lambda \to p\ell\bar{\nu}_\ell$ decays.

\subsection{Extraction of $|V_{us}|$}

To separate the theoretical contribution to the decay rate, we write
$\Gamma_{\Lambda \to p \ell\bar{\nu}_\ell}=
G_F^2|V_{us}|^2\,\tilde{\Gamma}^{\rm Theory}_{\Lambda \to p \ell\bar{\nu}_\ell},$
where $\tilde{\Gamma}$ depends only on the baryon masses and on the hadronic form factors, which can be computed from first principles in lattice QCD. The Fermi constant $G_F$ and the CKM matrix element $|V_{us}|$ instead enter as external inputs in the prediction of the decay rate.
Conversely, in light of the current tension in $|V_{us}|$, one may use experimental information on the decay rate to extract this parameter. This is achieved through the branching fraction
$\mathcal{B}_{\Lambda \to p \ell\bar{\nu}\ell}=
\tau_\Lambda\,
\Gamma_{\Lambda \to p \ell\bar{\nu}\ell},$
where $\tau_\Lambda$ denotes the $\Lambda$-hyperon lifetime. The CKM matrix element can then be determined as
\begin{equation}
|V_{us}|
=
\left(
\tau_\Lambda^{\rm Exp.}
\frac{\tilde{\Gamma}^{\rm Theory}_{\Lambda \to p \ell\bar{\nu}\ell}}
{\mathcal{B}^{\rm Exp.}_{\Lambda \to p \ell\bar{\nu}_\ell}}
\right)^{-1/2}.
\label{eq:vus}
\end{equation}

\subsection{Muon-to-electron decay-rate ratio}

A particularly clean observable is the ratio of muonic to electronic
decay rates,
\begin{equation}
R^{\mu e}
\equiv
\frac{\Gamma(\Lambda \to p\mu\bar{\nu}_\mu)}
     {\Gamma(\Lambda \to pe\bar{\nu}_e)}=\frac{\tilde{\Gamma}(\Lambda \to p\mu\bar{\nu}_\mu)}
     {\tilde{\Gamma}(\Lambda \to pe\bar{\nu}_e)}= \frac{\mathcal{B}(\Lambda \to p\mu\bar{\nu}_\mu)}
     {\mathcal{B}(\Lambda \to pe\bar{\nu}_e)}.
\label{eq:Rmue_def}
\end{equation}
This ratio is considered ``clean'' because it is independent of the CKM
matrix element, of the Fermi constant, and of the $\Lambda$ lifetime. Moreover, at next-to-leading
order in $\Delta = m_\Lambda - m_p$ it is also independent of hadronic form factors and is given by~\cite{Chang:2014iba}
\begin{equation}
R^{\mu e}_{\rm SM}
=
\sqrt{1-\frac{m_\mu^2}{\Delta^2}}
\left(1-\frac{9}{2}\frac{m_\mu^2}{\Delta^2}
      -4\frac{m_\mu^4}{\Delta^4}\right)
+\frac{15}{2}\frac{m_\mu^4}{\Delta^4}
\operatorname{artanh}
\left(\sqrt{1-\frac{m_\mu^2}{\Delta^2}}\right).
\label{eq.Rat}
\end{equation}
This observable allows to test lepton-flavor
universality and it is also sensitive to possible non-standard scalar or tensor
interactions \cite{Chang:2014iba,Liu:2023zvh}.

\section{Lattice QCD and Experimental input}
\subsection{Experimental inputs}
In order to evaluate Eq.~\eqref{eq:GammaMaster}, we use the charged-lepton masses
$m_e = 0.511\,\mathrm{MeV}$ and $m_\mu = 0.10566\,\mathrm{GeV}$, together with the
Fermi constant $G_F = 1.16638 \cdot (1 + 0.0105) \cdot 10^{-5}$\,GeV$^{-2}$\cite{ParticleDataGroup:2024cfk}, where the multiplicative factor $(1 + 0.0105)$ accounts for the radiative corrections~\cite{Garcia:1985xz}. For comparison with the lattice-QCD determination of the form-factor ratios, we use the experimental values listed in Table~\ref{tab:ExpCh}. In Eq. \eqref{eq:vus} we use the $\Lambda$ lifetime $\tau_\Lambda^\text{PDG} = 2.617(10) \cdot 10^{-10}$\,s \cite{ParticleDataGroup:2024cfk}. The currently available experimental branching fraction measurements are summarized in Table~\ref{tab:BrFr}.  Using these branching fractions, we also construct several experimental
determinations of the muon-to-electron decay-rate ratio defined in
Eq.~\eqref{eq:Rmue_def}, obtaining 
\begin{equation}
R^{\mu e}_{PDG} =  0.181(23)\,,\quad R^{\mu e}_{LHCb/PDG} =  0.175(12) \qq{and} R^{\mu e}_{PDG/BESIII} =  0.185(24).    
\end{equation}

\begin{table}[]
    \centering
    \small
    \begin{tabular}{|c|c|c|}
    \hline
        $g_1$/$f_1$ & $f_2$/$f_1$ & $g_2$/$f_1$ \\
        \hline
         $0.706^{+0.089}_{-0.086}$ \ \ (BESIII \cite{BESIII:2025hgj})& $0.77^{+0.53}_{-0.49}$ (BESIII \cite{BESIII:2025hgj})\ \ & $-0.19^{+0.65}_{-0.63}$ (BESIII \cite{BESIII:2025hgj})\\
          $0.178(15)$ \ \ \ \ (PDG \cite{ParticleDataGroup:2024cfk})& $0.15(30)$ (Dworkin \cite{PhysRevD.41.780}) & $-$\\
         \hline
    \end{tabular}
    \caption{Experimental form factor charge ratio. The values of \cite{BESIII:2025hgj} are the average between the particle and the anti-particle condributions.}
    \label{tab:ExpCh}
\end{table}

\begin{table}[]
    \centering
    \small
    \begin{tabular}{|ccc|ccc|}
    \hline
        &$\mathcal{B}(\Lambda \to pe\bar{\nu}_e) \cdot10^4$ & &&$\mathcal{B}(\Lambda \to p\mu\bar{\nu}_\mu) \cdot10^4$&  \\
        \hline
         &$8.16\pm 0.27$ &(BESIII \cite{BESIII:2025hgj})& &$1.46\pm0.10$ &(LHCb \cite{LHCb:2025wld})\\
         &$8.34\pm0.14$  &(PDG \cite{ParticleDataGroup:2024cfk})&& $1.51\pm0.19$ &(PDG \cite{ParticleDataGroup:2024cfk})\\
         \hline
    \end{tabular}
    \caption{Experimental branching fractions.}
    \label{tab:BrFr}
\end{table}
\subsection{Lattice QCD inputs}
The hadronic form factors entering the $\Lambda \to p$ semileptonic decay
are taken from our recent lattice-QCD calculation~\cite{Bacchio:2025auj}.
The calculation was performed on a single gauge ensemble generated by the
Extended Twisted Mass Collaboration with $N_f=2{+}1{+}1$ dynamical quark flavors,
corresponding to physical light, strange, and charm quark masses.
Working directly at the physical point eliminates the need for a chiral
extrapolation and constitutes a key advantage of the present study.

Matrix elements of the weak vector and axial-vector currents are extracted from
ratios of baryonic two- and three-point correlation functions constructed using
standard local interpolating fields for the proton and $\Lambda$ baryon.
Multiple source--sink separations are employed to control excited-state
contamination, and ground-state matrix elements are isolated using a
combination of the summation method and two-state fits.
Statistical uncertainties and correlations are estimated using bootstrap
resampling.

The momentum-transfer dependence of the form factors is parameterized using a
model-independent $z$-expansion, allowing for a controlled description over the
full kinematic range covered by the lattice data, including negative values of
$q^2$ and up to $q^2_{\max}$.
Correlations among all fit parameters are fully propagated to derived
quantities, such as decay rates and ratios of decay rates.
Further details of the lattice setup, renormalization procedure, and systematic
uncertainties can be found in Ref.~\cite{Bacchio:2025auj}.

\begin{figure}
\centering
{\includegraphics[width = 0.9\linewidth]{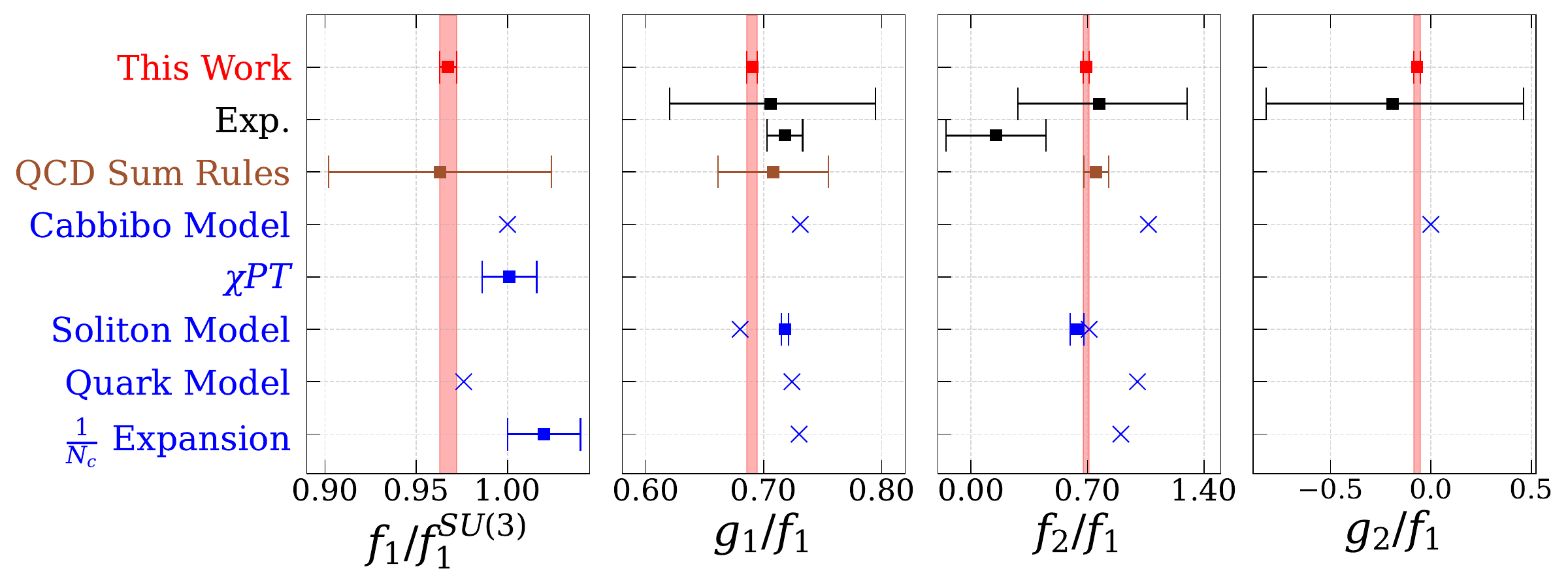}} 
\caption{Comparison of  ratios of coupling constants with experimental  and phenomenological ones. Points marked with a cross do not have uncertainties. The experimental values for $g_1/f_1$, $f_2/f_1$ and $g_2/f_1$ are taken from Table \ref{tab:ExpCh} and are shown in black. Results from phenomenology  include QCD sum rules~\cite{Zhang:2024ick,Zhang:2025qmg},  Cabibbo's model~\cite{Cabibbo:2003cu}, chiral perturbation theory ($\chi$PT)~\cite{Geng:2009ik}, the soliton model—where points with errors are from Ref.~\cite{Yang:2015era} and those without from Ref.~\cite{Ledwig:2008ku}—the quark model—where $f_1$ is taken from Ref.~\cite{Schlumpf:1994fb}, and $g_1/f_1$ and $f_2/f_1$ from Ref.~\cite{Faessler:2008ix}—and the $1/N_c$ expansion~\cite{Flores-Mendieta:1998tfv}.}
\label{fig:comp}
\end{figure}
\subsection{Baryon masses}
For the baryon masses entering the decay kinematics, we consider two distinct
approaches:

\paragraph{(a) Lattice-determined masses.}
Using the baryon masses determined on the lattice,
$m_N = 0.9471(28)\,\mathrm{GeV}$ and
$m_\Lambda = 1.1263(23)\,\mathrm{GeV}$, where
the resulting mass splitting
$m_\Lambda - m_N = 0.1792(24)\,\mathrm{GeV}$
carries a percent-level uncertainty.
Since the decay rate scales with the fifth power of this splitting at leading
order, the associated uncertainty is amplified by approximately a factor of
five and becomes the dominant source of error.
With this choice we obtain
\begin{equation}
    \tilde{\Gamma}_{\Lambda \to p e\bar{\nu}_e}
= 5.83(37)\, 10^{7}\,\mathrm{s}^{-1}\!,\quad\tilde{\Gamma}_{\Lambda \to p \mu\bar{\nu}_\mu}
= 1.01(12)\,10^{7}\,\mathrm{s}^{-1}\qq{and}R^{\mu e} = 0.1735(98)\,.
\end{equation}

\paragraph{(b) Experimental masses.}
Alternatively, using the experimental baryon masses
$m_p^{\mathrm{PDG}} = 0.93827\,$ $\mathrm{GeV}$ and
$m_\Lambda^{\mathrm{PDG}} = 1.11568\,\mathrm{GeV}$
eliminates the above uncertainty.
On the present ensemble, however, a percent-level discrepancy between lattice
and experimental baryon masses is observed, which is traceable to a cutoff effect that
vanishes in the continuum limit~\cite{Alexandrou:2023dlu}.
This choice therefore leads to a shift of approximately one standard deviation
in the decay rate relative to approach~(a) but much more precise results.
In this case we obtain
\begin{equation}
    \tilde{\Gamma}_{\Lambda \to p e\bar{\nu}_e}
= 5.546(51)\, 10^{7}\,\mathrm{s}^{-1}\!,\quad\tilde{\Gamma}_{\Lambda \to p \mu\bar{\nu}_\mu}
= 0.9228(82)\,10^{7}\,\mathrm{s}^{-1}\qq{and}R^{\mu e} = 0.16638(20)\,.
\end{equation}
Comparing the final results obtained with procedures (a) and (b) clearly highlights the dominant impact of the mass uncertainties. In a fully controlled lattice calculation, where all systematic effects are quantified, experimental masses can be employed to improve the overall precision. Conversely, when systematic uncertainties are not yet fully under control, lattice-determined masses may better reflect residual systematics. For this reason, in the following, we quote as our final result the more conservative determination obtained using the lattice masses.

\section{Results}

In Fig.~\ref{fig:comp}, we compare commonly used ratios of coupling constants
with experimental data and phenomenological determinations.
We observe good agreement with recent experimental results from BESIII and with
QCD sum-rule calculations, while achieving a significantly higher precision.
Our lattice-QCD results are
\begin{equation}
\frac{f_1}{f_1^{\mathrm{SU(3)}}}=0.9674(47),\,\quad
\frac{g_1}{f_1}=0.6902(44)\,,\quad
\frac{f_2}{f_1}=0.693(17)\,,\quad
\frac{g_2}{f_1}=-0.069(16)\,.
\end{equation}
Here $f_1^{\mathrm{SU(3)}}{=}\sqrt{3/2}$ denotes the SU(3)-flavor symmetry limit
predicted by the Ademollo--Gatto theorem.
\begin{figure}
\centering
{\includegraphics[width = 0.8\linewidth]{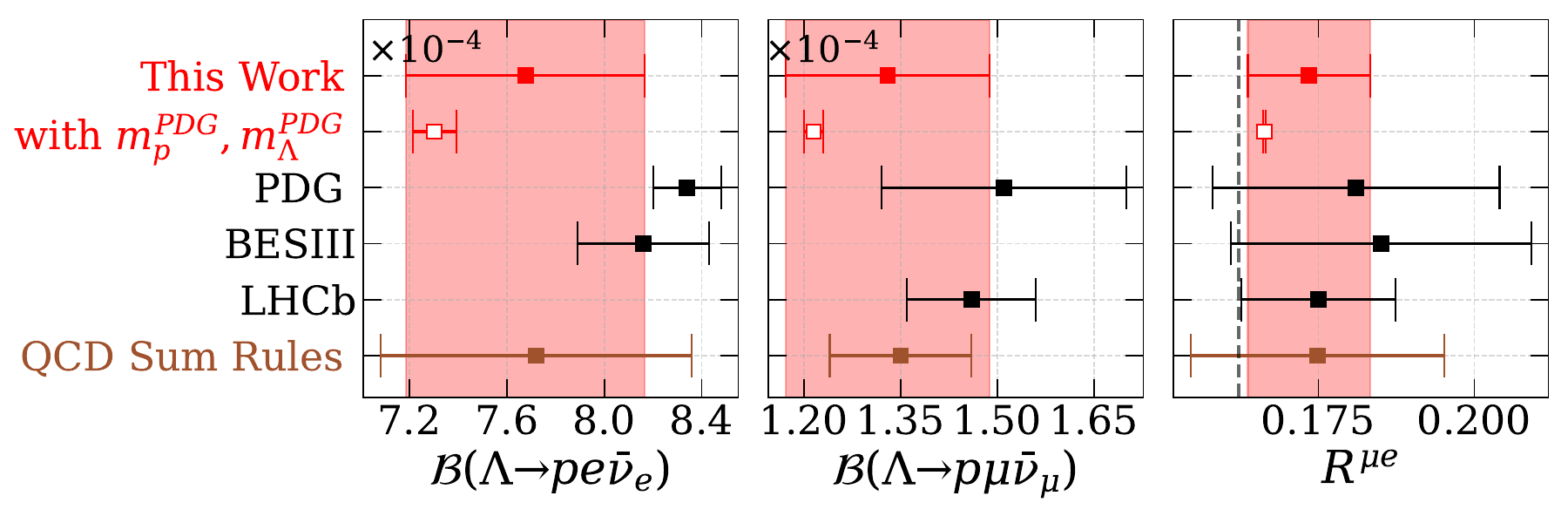}} 
\caption{Results on the branching fractions in the electron and muon channels and their ratio. For this work we also provide the branching fraction using as input the PDG average value $|V_{us}^{\rm PDG}| = 0.2243(8)$. The Experimental values $R^{\mu e}_{PDG}$ =  0.181(23), $R^{\mu e}_{PDG/BESIII}$ =  0.185(24) and $R^{\mu e}_{LHCb/PDG}$ =  0.175(12) are shown by the black points, and the QCD sum rules value by the brown points~\cite{Zhang:2024ick,Zhang:2025qmg}. The grey dashed line represents the 0.162 value which comes from Eq. \eqref{eq.Rat} when the experimental values are adopted. 
}
\label{fig:decrarte}
\end{figure}

\begin{figure}
\centering
{\includegraphics[width = 0.7\linewidth]{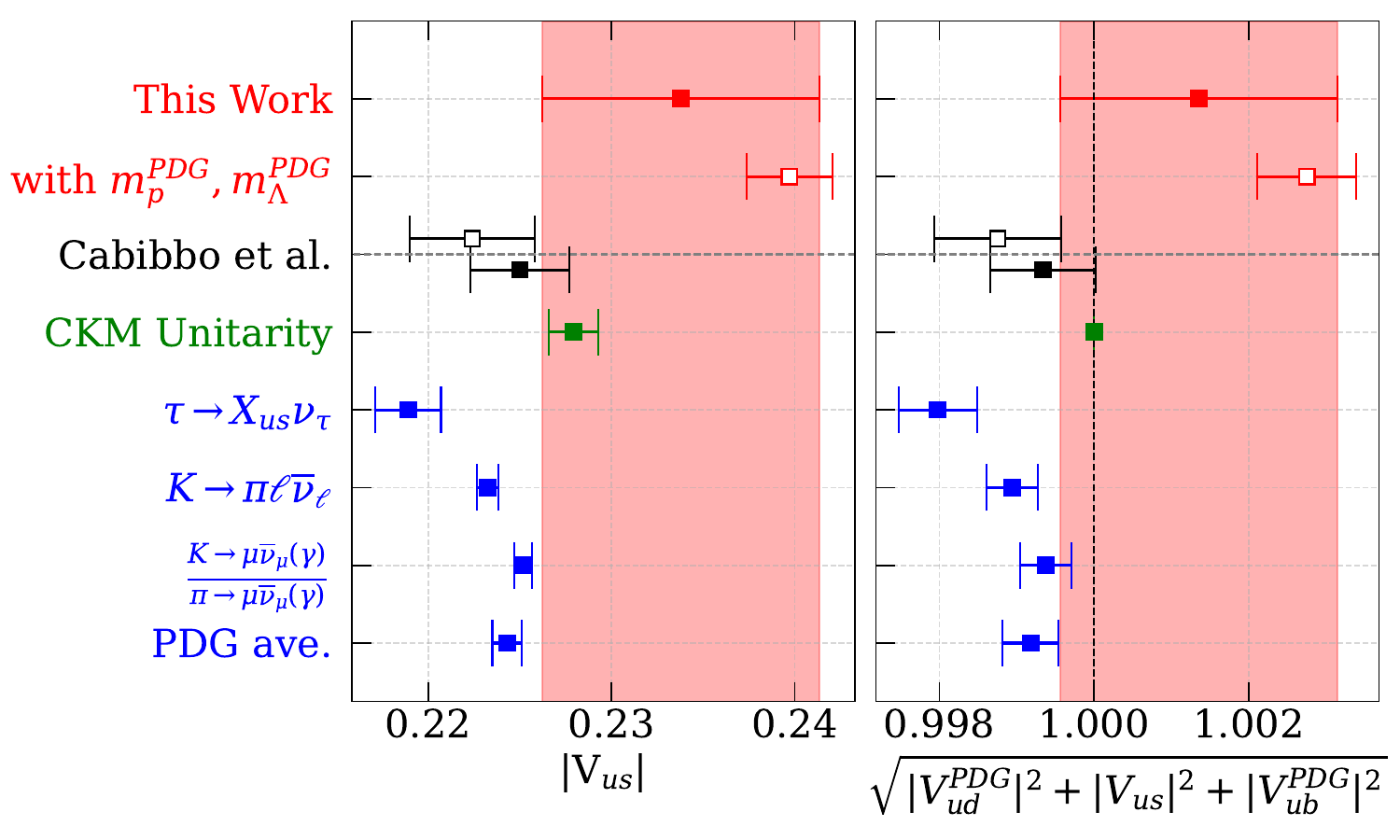}} 
\caption{Determination of $V_{us}$ (left) and the resulting CKM unitarity relation (right). Our results are shown in red: the top one uses lattice-determined nucleon and $\Lambda$ masses, approach (a), while the second one with open symbols uses PDG values, approach (b). The  black points correspond to the values by Cabbibo \textit{et al.}~\cite{Cabibbo:2003cu}, where the empty point is the value derived from the $\Lambda$ semileptonic decay, while the full point combines results from various hyperon decays. The green point is obtained from the unitarity relation. The remaining blue points show results on decay rates using lattice QCD combined with experimental data. Going from top to bottom, the  first blue square is from inclusive $\tau$ decays~\cite{ExtendedTwistedMass:2024myu}, the second from kaon semileptonic decays~\cite{Moulson:2017ive,Carrasco:2016kpy,Bazavov:2018kjg,FlavourLatticeAveragingGroupFLAG:2024oxs}, the third from the ratio of kaon to pion leptonic decays in the muonic channel~\cite{Dowdall:2013rya,Carrasco:2014poa,Bazavov:2017lyh,Miller:2020xhy,Alexandrou:2021bfr,Moulson:2017ive,FlavourLatticeAveragingGroupFLAG:2024oxs}, and the last is the PDG average of the latter two~\cite{ParticleDataGroup:2024cfk}. 
}
\label{fig:vues}
\end{figure}
A visualization of the branching fractions and the muon-to-electron decay-rate ratios is shown in
Fig.~\ref{fig:decrarte}.
Our results are consistent with the available experimental determinations and
with QCD sum-rule predictions.
In particular, the value obtained when using experimental baryon masses
(case~(b)) highlights the significant impact of baryon-mass uncertainties in the
lattice-QCD calculation on the final results.
All ratios lie above the value $R^{\mu e}=0.162$, corresponding to the NLO
perturbative estimate.
Comparing our result from case~(b) with this value suggests that the observed
difference is primarily due to the fully integrated treatment of the decay rate
rather than to effects beyond the Standard Model.

Finally, we extract the value of $|V_{us}|$ using the electron channel PDG experimental branching
fraction listed in Table~\ref{tab:BrFr}.
The results are summarized in Fig.~\ref{fig:vues}, where we also test the
unitarity of the first row of the CKM matrix, which is expected to equal unity in
the Standard Model.
For this purpose, we use
$|V_{ud}^{\mathrm{PDG}}| = 0.97367(32)$ and
$|V_{ub}^{\mathrm{PDG}}| = 3.82(20)\times 10^{-3}$~\cite{ParticleDataGroup:2024cfk}.
For the case where the lattice
baryon masses (case~(a)) are implemented, we obtain
\begin{align}
|V_{us}| &=
\left(
\frac{\tau_\Lambda^{\mathrm{PDG}}\,\tilde{\Gamma}^{\text{case (a)}}_{\Lambda \to p \ell\bar{\nu}_\ell}}
     {\mathcal{B}(\Lambda \to p e\bar{\nu}_e)^{\mathrm{PDG}}}
\right)^{-1/2}
= 0.2338(75),
\end{align}
leading to
\begin{align}
\sqrt{|V_{ud}^{\mathrm{PDG}}|^2 + |V_{us}|^2 + |V_{ub}^{\mathrm{PDG}}|^2}
= 1.0014(18).
\end{align}
Within the current uncertainties, first-row CKM unitarity is therefore
satisfied, although the precision of the test is limited by the uncertainty on
$|V_{us}|$. Table \ref{tab:vus} gives the $|V_{us}|$ CKM matrix element value calculated with the experimental branching fractions from Table \ref{tab:BrFr}, where cases (a) and (b) are implemented.

\begin{table}[]
    \centering
    \small
\begin{tabular}{|c|c|c|c|c|}
       \hline$|V_{us}|$&$\mathcal{B}^{PDG}_{(\Lambda \to pe\bar{\nu}_e)} $ &$\mathcal{B}^{BESIII}_{(\Lambda \to pe\bar{\nu}_e)} $&$\mathcal{B}^{PDG}_{(\Lambda \to p\mu\bar{\nu}_\mu)} $&$\mathcal{B}^{LHCb}_{(\Lambda \to p\mu\bar{\nu}_\mu)} $ \\
       \hline
       case (a) &0.2338(77)&0.2313(83)&0.239(21)&0.235(16)  \\
        case (b) &0.2397(23) &0.2371(41)&0.250(16)&0.2459(85)\\
        \hline
    \end{tabular}
    \caption{$|V_{us}|$ CKM matrix element calculated with the experimental branching fractions from Table \ref{tab:BrFr}.}
    \label{tab:vus}
\end{table}
\section{Conclusion}

In this work we have combined recent experimental measurements of $\Lambda \to p\ell\bar{\nu}_\ell$ decays with lattice-QCD form factors to investigate their phenomenological implications and to extract the CKM matrix element $|V_{us}|$. Our study shows that lattice QCD already provides a consistent and reliable description of the relevant hadronic matrix elements at a single lattice spacing.

We further find that sub-percent precision can in principle be achieved for several observables, including effective couplings and the muon-to-electron decay-rate ratio. Achieving this level of accuracy, however, requires full control of systematic effects, in particular the continuum limit and a precise tuning to the isosymmetric QCD point. In our final results we therefore adopt baryon masses determined on the lattice, which dominate the present uncertainty and effectively account for residual systematics. This choice reflects the strong sensitivity of the phase-space factors to small distortions of the baryon spectrum at finite lattice spacing.

These findings motivate extending the present calculation to multiple lattice spacings in order to perform a controlled continuum extrapolation of the form factors. Reaching the continuum limit will allow the consistent use of experimental baryon masses in the decay kinematics and is expected to significantly reduce the uncertainty on $|V_{us}|$, limited  then by the experimental uncertainty on the branching fraction, as shown in Fig.~\ref{fig:vues}.
With improved lattice inputs and forthcoming high-precision experimental measurements, semileptonic hyperon decays have the potential to become a competitive and fully independent probe of CKM unitarity.

\section*{\small ACKNOWLEDGMENTS}
\small
We thank Constantia Alexandrou for her insightful comments and mentorship as A.K.'s PhD advisor. We also thank the members of the ETM Collaboration for their constructive cooperation. S.B. acknowledge support from \texttt{POST-DOC/0524/0001} and \texttt{VISION ERC - PATH 2/0524/0001}, co-financed by the European Regional Development Fund and the Republic of Cyprus through the Research and Innovation Foundation within the framework of the Cohesion Policy Programme “THALIA 2021-2027”. A.K. acknowledges financial support from ``The three-dimensional structure of the nucleon from lattice QCD'' \texttt{3D-N-LQCD} program, funded by the University of Cyprus. We gratefully acknowledge the Gauss Centre for Supercomputing e.V. (www.gauss-centre.eu) for computing time on JUWELS Booster at the Jülich Supercomputing Centre (JSC). We also acknowledge the Swiss National Supercomputing Centre (CSCS) and the EuroHPC Joint Undertaking for access to the Daint-Alps supercomputer. We are grateful to CINECA and the EuroHPC JU for access to the supercomputing facilities hosted at CINECA and Leonardo-Booster.

\begingroup
\footnotesize
\setlength{\itemsep}{0pt}
\setlength{\parskip}{0pt}
\bibliography{main}
\endgroup
\end{document}